\documentclass[letter]{aa}     

\usepackage{multirow}
\usepackage{makecell}
\usepackage{graphicx}
\usepackage{txfonts}
\usepackage{lipsum}
\usepackage{subcaption}         
\usepackage{lscape}             
\usepackage{placeins}           

\usepackage{graphicx, array, booktabs, tikz}
\usetikzlibrary{decorations.pathmorphing, arrows.meta, angles, quotes}
\usepackage[colorlinks=true,allcolors=blue]{hyperref}

\begin{document} 


\title{Ultra-long MeV transient from a relativistic jet: a tidal disruption event candidate}

\author{
Gor~Oganesyan\inst{1,2}\thanks{\email{gor.oganesyan@gssi.it}}
\and
Elias~Kammoun\inst{3}\thanks{\email{ekammoun@caltech.edu}}
\and
Annarita~Ierardi\inst{1,2}\thanks{\email{annarita.ierardi@gssi.it}}
\and
Alessio~Ludovico~De~Santis\inst{1,2}
\and 
Biswajit~Banerjee\inst{1,2}
\and 
Emanuele~Sobacchi\inst{1,2}
\and 
Felix~Aharonian\inst{4,5,6}
\and 
Samanta~Macera\inst{1,2}
\and
Pawan~Tiwari\inst{1,2}
\and
Alessio~Mei\inst{1,2}
\and 
Shraddha~Mohnani\inst{7}
\and 
Stefano~Ascenzi\inst{1,2}
\and
Samuele~Ronchini\inst{8,9}
\and
Marica~Branchesi\inst{1,2}
}

\institute{
Gran Sasso Science Institute, Viale F. Crispi 7, L'Aquila (AQ), I-67100, Italy
\and 
INFN - Laboratori Nazionali del Gran Sasso, L'Aquila (AQ), I-67100, Italy
\and 
Cahill Center for Astronomy \& Astrophysics, California Institute of Technology, Pasadena, CA 91125, USA
\and 
Dublin Institute for Advanced Studies, 31 Fitzwilliam Place, Dublin 2, Ireland
\and 
Max-Planck-Institut f\"ur Kernphysik, Saupfercheckweg 1, 69117 Heidelberg, Germany
\and 
Yerevan State University, 1 Alek Manukyan St, Yerevan 0025, Armenia
\and 
Department of Astronomy, Astrophysics and Space Engineering Indian Institute of Technology Indore, Simrol, Khandwa Road, Indore 453552, Madhya Pradesh, India
\and 
Department of Astronomy and Astrophysics, The Pennsylvania State University, 525 Davey Lab, University Park, PA 16802, USA
\and Institute for Gravitation \& the Cosmos, The Pennsylvania State University, University Park, PA 16802, USA
}

\date{Received xxx; accepted xxx}

\abstract{
On July 2, 2025, the Gamma-ray Burst Monitor (GBM) onboard the \textit{Fermi} Gamma-ray space telescope detected three short-duration MeV transients with overlapping sky locations. These events, named as GRB\,250702D, B, and E (collectively referred to as DBE), triggered the detector with delays of approximately $1-2$\,hours between each burst. Follow-up observations of this unusually long MeV transient (lasting $>$3\,hours) by the \textit{Neil Gehrels Swift Observatory} and the \textit{Nuclear Spectroscopic Telescope Array} over a period of $\sim$10\,days revealed a steep temporal decline in soft X-rays ($\propto t^{-1.9 \pm 0.1}$). The time-averaged spectra during the outbursts are well described by a single power law $dN_{\gamma}/dE \propto E^{-1.5}$, while upper limits above 100 MeV imply a spectral cutoff between 10 MeV and 100 MeV. Using standard $\gamma$-ray transparency arguments, we derive a lower limit on the bulk Lorentz factor. Combined with the steep decline in X-rays, these constraints point to a relativistic jet origin. The properties of DBE are inconsistent with established GRB spectral–energy correlations, disfavoring classical long-GRB progenitors. Instead, the basic characteristics of DBE resemble those of previously reported jetted tidal disruption events (TDEs), though alternative progenitor channels cannot be excluded. In the relativistic TDE scenario, DBE is the first one with detected MeV $\gamma$-ray emission. We argue that the observed emission is most likely produced by synchrotron radiation from sub-TeV electrons.}

\keywords{
radiation mechanisms: non-thermal --
acceleration of particles --
gamma rays: general --
black hole physics --
relativistic processes
}

\maketitle
%
\section{Introduction}
The Gamma-ray Burst Monitor (GBM) onboard the \textit{Fermi} Gamma-ray Space Telescope detected and localized five Gamma-Ray Bursts (GRBs\footnote{Not all short-lasting MeV transients named as GRBs upon their trigger originate from classical GRB progenitors, such as collapsars and binary neutron star mergers.}) on July 2, 2025. Three of these MeV transients, namely GRB\,250702D, B, and E (hereafter DBE), have overlapping sky positions, likely originating from the same source \citep{DBE}. The detection of hard X-ray emission from DBE was also reported by Konus-WIND \citep{Konus}, Swift-BAT/GUANO \citep{guano} and the Space-based multi-band astronomical Variable Objects Monitor \citep[SVOM;][]{SVOM}. Soft X-ray detections spatially coincident with DBE were reported by the Einstein Probe \citep[EP\,250702A;][]{EP} and the Monitor of All-sky X-ray Image \citep{MAXI}, in observations preceding the first \textit{Fermi}/GBM trigger.

Follow-up observations of EP\,250702A/DBE by the X-Ray Telescope (XRT) onboard the \textit{Neil Gehrels Swift Observatory} (\textit{Swift}, hereafter) revealed an associated fast-declining soft X-ray transient \citep[$\propto t^{-2}$;][]{Swift}. Observations by the \textit{Nuclear Spectroscopic Telescope Array} (\textit{NuSTAR}) in hard X-rays ($3-79\,\rm keV$), performed $\sim 1$\,day after the trigger of DBE, were consistent with the steep decline in soft X-rays \citep{O'Connor2025}.  A similar rapid decline was observed in the near-infrared \citep[NIR;][]{Levan2025}.

The duration of DBE in the MeV phase ($\rm >3\,hours$) significantly exceeds the typical timescales of classical GRBs ($0.1-100\,\rm s$). Its X-ray counterpart also evolves faster than typical GRB afterglows, where the luminosity decreases as $\propto t^{-1}$ before the jet break time \citep{Meszaros1997}, i.e. when $1/\Gamma << \theta_{j}$, where $\Gamma$ is the bulk Lorentz factor and $\theta_j$ is the opening angle of the jet. These two basic observables (i.e., the prolonged MeV $\gamma$-ray emission and the steep X-ray decline) make DBE a candidate for a relativistic Tidal Disruption Event \citep[TDE;][]{Levan2025}. The presence of hours-long outbursts preceding the classical steep decline phase suggests the formation of a relativistic jet. To date, only four relativistic TDEs have been identified: Swift\,J164449.3+573451 \citep{Burrows2011,Bloom2011,Levan2011}, Swift\,J2058.4+0516 \citep{Cenko2012,Pasham2015}, Swift\,J1112.2-8238 \citep{Brown2015}, and AT\,2022cmc \citep{Andreoni2022,Pasham2023}. These events were detected up to hard X-rays ($<200$\,keV), but not in MeV $\gamma$-rays.

In this Letter, we investigate the high-energy radiation associated with DBE, from soft X-rays to $\gamma$-rays. We analyze each DBE outburst using publicly available data from \textit{Fermi}/GBM ($\rm 8\,keV - 40\,MeV$) and \textit{Fermi}/Large Area Telescope (LAT, $> 100\rm \,MeV$). We also analyze the soft and hard X-ray emission of DBE in the first 10 days, provided by publicly available \textit{Swift}/XRT observations, and a dedicated \textit{NuSTAR} Director Discretionary Time (DDT) observation obtained by our team. We then briefly discuss the properties of the emission region, where the non-thermal $\gamma$-ray photons are produced.   

\section{Temporal profile}
\label{section:lightcurves}

\begin{figure}
\includegraphics[width=\linewidth]{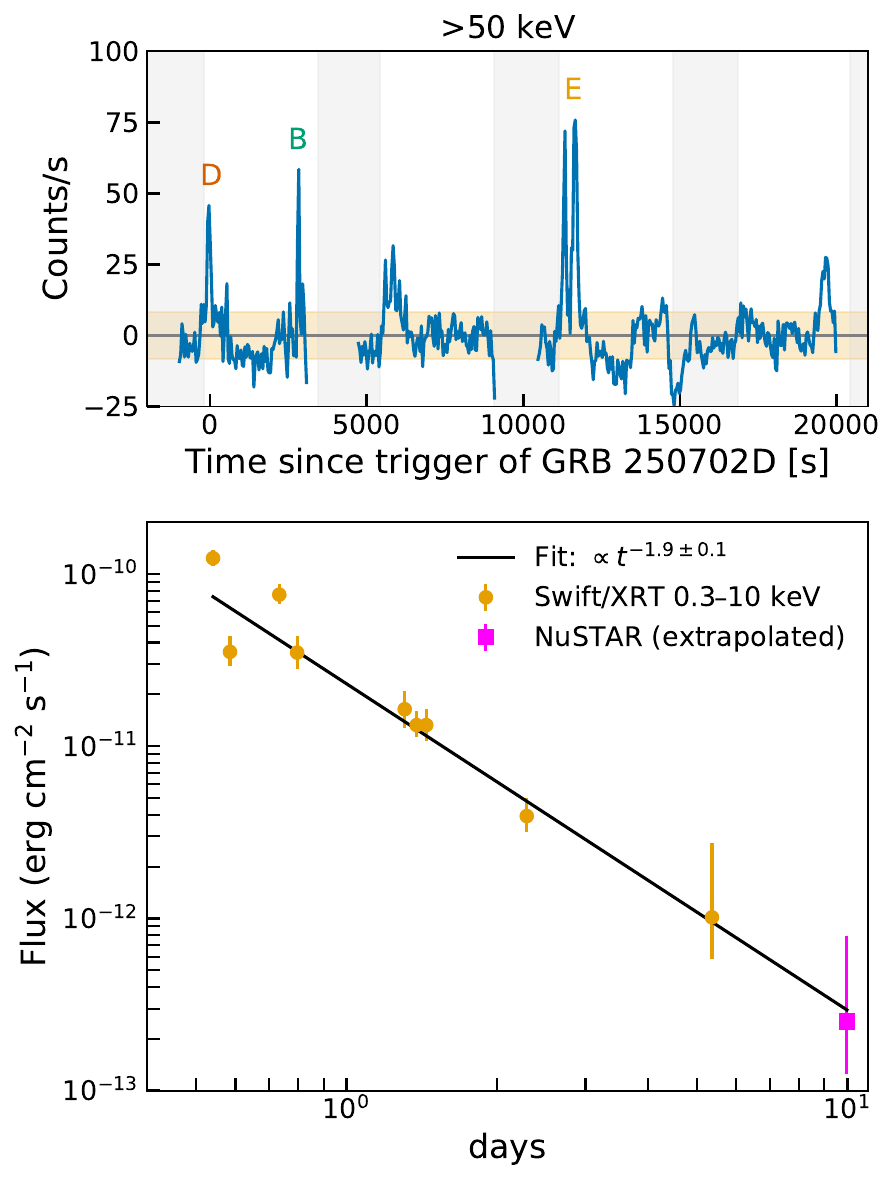}
\caption{Upper panel: Background-subtracted $\rm 5\,s$-binned light curve of the hard X-rays/MeV emission detected by \textit{Fermi}/GBM ($50-900$\,keV). Gray shaded regions indicate time bins during which the source was occulted by the Earth. The yellow shaded region marks $1 \sigma$ noise level in the light curve. Lower panel: Light curve of the soft X-ray counterpart obtained from time-resolved spectral analysis of \textit{Swift}/XRT (0.3–10 keV, yellow circles). The flux at $0.3-10$\,keV at 10 days (pink square) was estimated by extrapolating the spectral model fitted to the \textit{NuSTAR} data (fitted in the $3-30$\,keV range).}
\label{fig:LCs}
\end{figure}

We extract the background-subtracted MeV light curve over a 5.5-hour interval following the trigger of 250702D (see the upper panel of Fig.\,\ref{fig:LCs} and Appendix\,\ref{OSV}). The first two events, 250702D and 250702B, have similar durations of $\sim100-150$\,s and are separated by a quiescent period of $\sim 45$ minutes. A third associated burst, 250702E, occurred $\sim$3\,hours later and lasted $\sim 600$\,s. Large parts of the intermediate temporal windows between these three events were not accessible due to Earth's occultation. A weak emission episode at\footnote{$t_0$ is defined as the trigger time of 250702D, on 2025-07-02 13:09:02.03\,(UT).} $\sim t_0 + 6000\,\rm s$ originates from GRB\,250702C \citep{C}, which is not related to DBE \citep{DBE}.

We performed a time-resolved spectral analysis of the soft X-ray emission observed by \textit{Swift}/XRT ($0.3-10$\,keV) and \textit{NuSTAR} \citep[][]{Harrison13} over the period from 0.5 to 10 days post-trigger of 250702D. The details of the spectral analysis are presented in Appendix~\ref{Xrays}. The X-ray light curve shows a steep temporal decline, following $t^{-1.9 \pm 0.1}$ (lower panel of Fig.\,\ref{fig:LCs}). 

\section{$\gamma$-ray spectra}
\label{section:gamma-ray}

\begin{figure*} 
  \centering
  \includegraphics[width=0.65\textwidth]{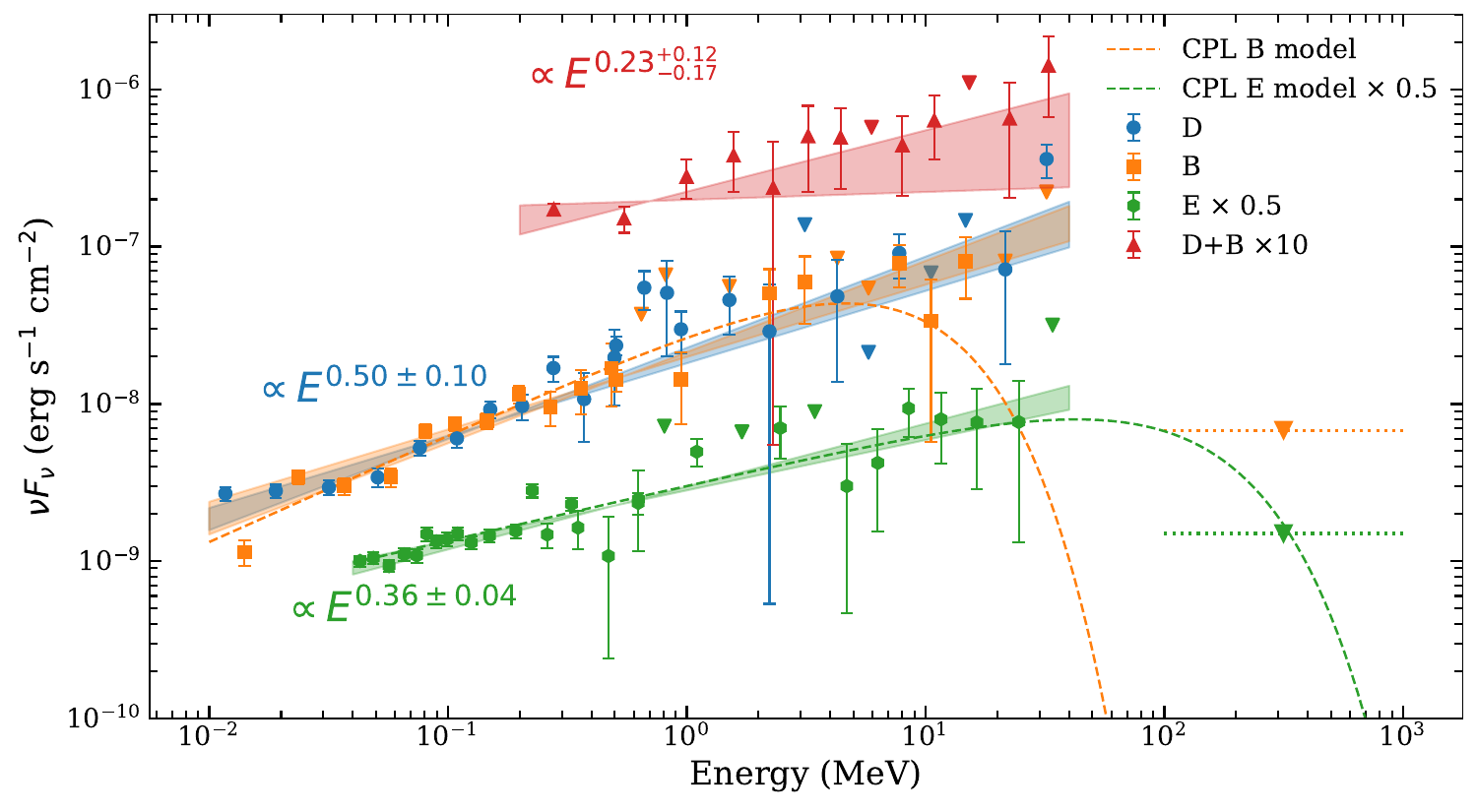}
  \caption{Spectral energy distributions of events D, B, and E. The derived data points for the time-integrated spectra of D, B, and E, along with the best-fit power-law models (with $1\sigma$ uncertainties) are shown using the blue, orange, and green symbols and shaded regions, respectively. The dashed orange and green lines represent average cutoff power-law models for the time-integrated spectra of events B and E. Upper limits between 100\,MeV and 1\,GeV ($3\sigma$) are shown as inverted triangles. The combined D and B spectrum above 200\,keV, along with its best-fit model ($1\sigma$), is shown using red triangles and a red shaded region. The corresponding spectral slopes ($2-\alpha$) are also indicated.}
  \label{fig:spectra}
\end{figure*}

We extract the \textit{Fermi}/GBM spectra during the three emission episodes (D, B, and E) from the daily GBM data. The background spectra for the short episodes (D and B) are estimated using standard polynomial fits to time intervals before and after each burst. To evaluate the background during the long-duration episode E, we adopt the average orbital method. Details of GBM data extraction and analysis are provided in Appendix\,\ref{GBMspectra}. 

We first model the spectra of episodes D, B, and E using a simple power-law (PL) model, $dN/dE \propto E^{-\alpha}$ (Fig.~\ref{fig:spectra}). The best fit to the data in the energy range between 10\,keV (40\,keV for E) and 40\,MeV provides $\alpha = 1.5 \pm 0.1$ for D and B, and a slightly softer spectrum for E ($\alpha = 1.64 \pm 0.04$).  

We then test for the presence of a high-energy softening by fitting the spectra with a cutoff power-law (CPL) model. We find no significant improvement in the fit compared to the simple PL model. The cutoff energy is poorly constrained in all cases: $E_{\rm cut} > 4$\,MeV for D, $E_{\rm cut} = 7_{-6}^{+13}$\,MeV for B, and $E_{\rm cut} > 14$\,MeV for E. We compare the best-fit CPL models with the $3\sigma$ upper limits from \textit{Fermi}/LAT at $0.1-1$\,GeV (see Appendix~\ref{LAT}). These upper limits suggest that the cutoff energy should be between the GBM and LAT energy bands, i.e.~10 MeV$<E_{\rm cut}<$ 100 MeV. Our analysis shows that the time-resolved and time-integrated spectra of DBE are incompatible with the established spectral–energy correlations (Amati and Yonetoku relations) for long-duration GRBs (Appendix~\ref{correlations}).

The spectra of episodes D and B are consistent in normalization and spectral shape. Above 200\,keV, the background is well characterized by the average orbital method. Thus, we extract the cumulative D and B spectrum. Fitting it with a PL model returns $\alpha = 1.77_{-0.12}^{+0.17}$, indicating a possible softening at high energies.  

\section{Discussion}
\label{section:discussion}

The particularly long duration ($>3$\,hours) and variable MeV radiation is followed by a steep temporal decline in X-rays. The spectra in the MeV $\gamma$-rays are markedly different from those of classical GRBs. We find $\alpha \sim 1.5$ or softer in DBE, as opposed to $\alpha \sim 1$ in GRBs \citep{Band1993}. Moreover, the DBE spectra extend beyond $\sim 0.1-1$\,MeV, which is unusual for both short and long GRBs. The inferred photon index, $\alpha \sim 1.5$, can be produced by a cooled population of electrons with a distribution $dN_e/d\gamma_e \propto \gamma_e^{-2}$. In Appendix~\ref{physics}, we argue that the observed spectra are most likely produced by synchrotron radiation from sub-TeV electrons.

Provided the particularly long MeV duration and their spectral shape, together with the steep X-ray decline ($\rm 0.5-10$\,days), we disfavor the origin via classical channels of formation of GRBs. During the tidal disruption of a star by a massive compact object, the fallback of stellar debris typically produces a flux decay $\propto t^{-5/3}$ \citep{Rees1988}. However, the X-ray spectra of DBE events are non-thermal, with photon indices between 1.4 and 1.6. The rapid X-ray decline may therefore be interpreted as afterglow emission from a forward shock once $1/\Gamma \gtrsim \theta_j$ \citep{Rhoads1997}, or alternatively as emission from a jet with diminishing power.

Several arguments were put forward supporting the extragalactic origin of DBE \citep{Levan2025}, including its association with a candidate host galaxy of complex morphology. A transient relativistic jet formed as a consequence of the tidal disruption of a star by a compact object is a good candidate for long-duration MeV radiation. The previous four relativistic TDEs lack the emission in MeV $\gamma$-rays and were observed only up to hard X-rays ($<200$\,keV). The absence of associated MeV radiation can be referred to the poor sensitivity of MeV instruments. In Appendix~\ref{comparison}, we compare the energetics as well as the temporal and spectral properties of DBE with those of previously reported relativistic TDE candidates. Although the sample size is limited, the X-ray variability, spectral shape, and energetics of DBE are consistent with those of the other candidates.

The individual outburst rise time and duration are linked to the density profile of the disrupted star and the mass of the compact object. It was previously argued that a variability time-scale of $\sim$ 1000 s implies that the disrupted star is most probably a white dwarf \citep{Krolik2011}. The disruption of a white dwarf requires that the mass of the compact object is $< 10^{5} M_{\odot}$. In the MeV $\gamma$-rays, DBE has even shorter observed variability of $\sim 100-600$\,s, which indicates that the compact object is an intermediate mass black hole \citep{Levan2025}.

\subsection{Constraints on the bulk Lorentz factor}

We estimate a lower limit on the bulk Lorentz factor of the jet, assuming that the DBE emission is subject to the usual MeV compactness constraint\footnote{Similar constraints were applied to classical GRBs in \citet{Ruderman1975,Schmidt1978,Lithwick2001}. See \citet{Ravasio2024} for more detailed derivations.}:
\begin{equation}
\Gamma >
\left[
\frac{
\eta(\beta) \sigma_T \left( \frac{\epsilon_{\max}}{2} \right)^{2\beta - 1} (1+z)^{2\beta} d_L^2(z) F_0(1 - \beta)
}{
2 \delta t_v m_e c^4 \left[ 1 - \left( \frac{\epsilon_0}{\epsilon_{\max}} \right)^{1 - \beta} \right]
}
\right]^{\frac{1}{4 + 2\beta}},
\end{equation}
where $\beta\sim 0.5$ is the spectral index ($F_{\nu} \propto E^{-\beta}$, where $\beta = \alpha-1$), $\eta(\beta)=7/6(2+\beta)(1+\beta)^{5/3}$ \citep{Svensson1987}, $\delta t_{v}\sim 100$\,s is the observed variability time-scale, $\epsilon_{0}\sim 0.02$ and $\epsilon_{\max}\sim 78$ are the minimum and maximum energies of the observed PL spectrum in units of the electron rest mass, $F_{0}\sim {3 \times 10^{-7}}{\rm\; erg}{\rm\; cm}^{-2}{\rm\; s}^{-1}$ is the flux integrated between $\epsilon_{0}$ and $\epsilon_{\max}$, $z$ and $d_{L}$ are the redshift and luminosity distance to the source, respectively. 

\begin{figure}
\includegraphics[width=0.90\linewidth]{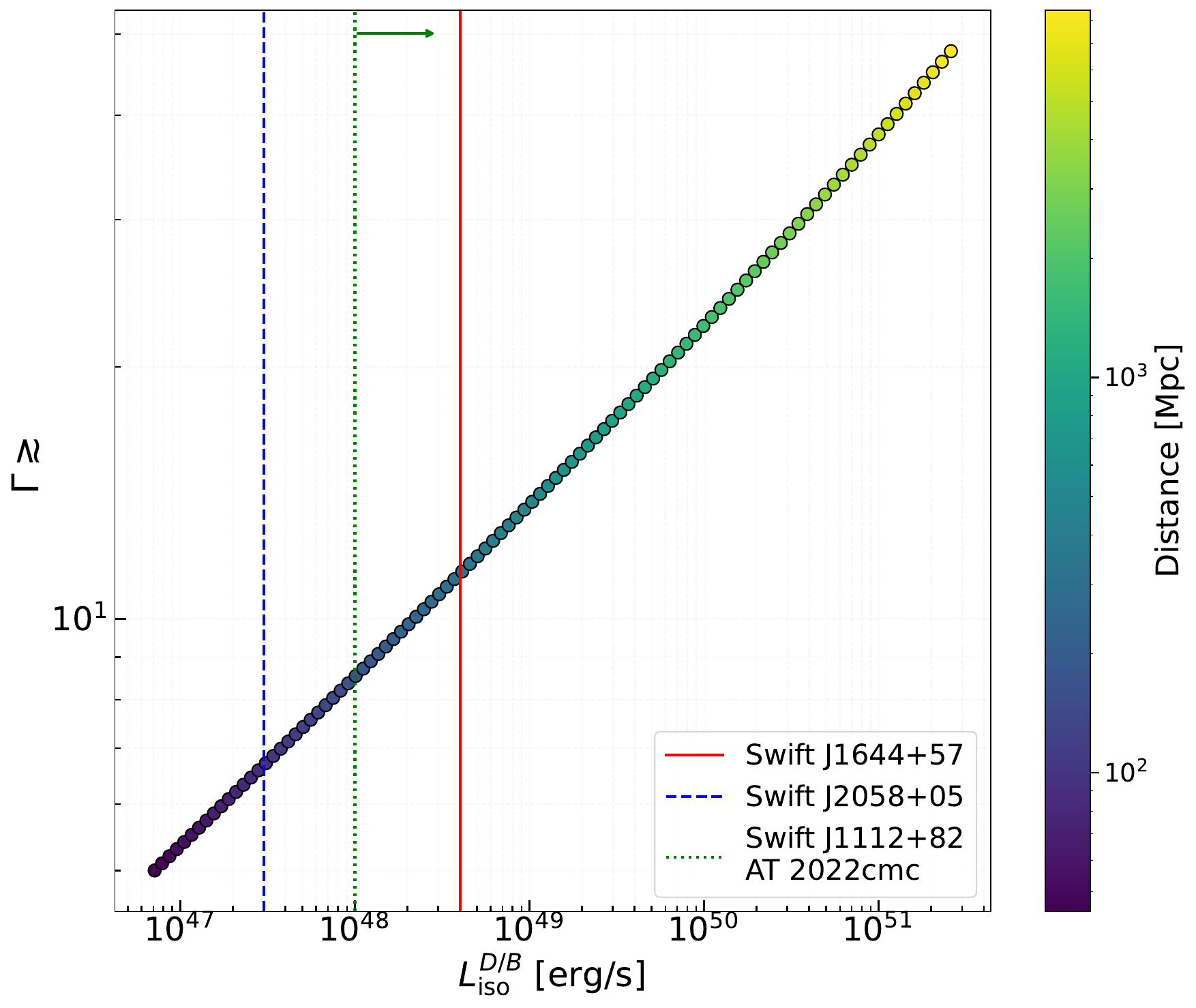}
\caption{Lower limit on the bulk Lorentz factor of the jet as a function of the isotropic-equivalent luminosity of the D and B emission episodes.
The color bar indicates the corresponding luminosity distance to DBE.
For comparison, we show the luminosity (brightest hard X-ray outbursts) measured for Swift J2058+05 \citep{Cenko2012,Pasham2015}, Swift\,J1112+82 \citep{Brown2015}, AT\,2022cmc \citep{Andreoni2022,Pasham2023}, and Swift\,J1644+57 \citep{Burrows2011} as dashed blue, dotted green, and solid red vertical lines.}
    \label{fig:Gamma}
\end{figure}

In Fig.~\ref{fig:Gamma}, we show the lower limit on the bulk Lorentz factor of the jet, $\Gamma > 7 d_{L, \rm 100 \ Mpc}^{2/5}$, as a function of the isotropic-equivalent outburst luminosity, $L_{\rm iso} \sim 4 \times 10^{47} d_{L, \rm 100 \ Mpc}^{2} {\rm erg}\, {\rm s}^{-1}$, and source distance. We find that for $\Gamma > 10$, DBE is placed in the same category as the four previously reported jetted TDEs in terms of outburst luminosity.

The total fluence emitted during the three episodes across $\epsilon_{0}(\sim 0.02) - \epsilon_{\max}(\sim 78)$ is $\sim 10^{-4} \ \rm erg \ cm^{-2}$. Given the extension of the DBE spectra up to multi-MeV $\gamma$-rays ($>$ 1 MeV), the MeV emission provides a reasonable approximation of the total radiative energy budget. In Fig.~\ref{fig:Gamma2}, we show $\Gamma$ as a function of isotropic-equivalent energy. To match the energetics of the previously reported jetted TDEs, a value of $\Gamma \sim 30$ is required. Combining our constraints on the luminosity of individual outbursts and the total energy budget, we conclude that $10 < \Gamma < 30$ and $100 {\rm\, Mpc} < d_{L} < 2 {\rm\, Gpc}$, consistent with the redshift constraints from the host galaxy  \citep[$z<1$;][]{Levan2025}.\footnote{In these estimates, we assume the jet is viewed on-axis. However, some TDE jets may be misaligned to account for late-time radio afterglow observations \citep{Beniamini2023}. If the DBE jet is misaligned, then the true Lorentz factor could be even larger.}

\section{Conclusions}

We have presented a temporal and spectral analysis of GRB\,250702D/B/E (DBE), three MeV transients with durations of $100-600$\,s, separated by $\sim1-3$\,hours. The spatial coincidence of these events indicates a common origin, suggesting ultra-long central engine activity lasting more than 3 hours. Our findings can be summarized as follows:

\begin{itemize}
    \item The prolonged MeV emission and the subsequent steep decline in the X-ray light curve in the first 10 days are different from classical GRBs, pointing instead to a TDE involving a relativistic jet.
    
    \item All three $\gamma$-ray episodes exhibit soft power-law spectra with indices $\alpha$ ($dN/dE \propto E^{-\alpha}$) ranging between 1.5 and 1.7 over the $\rm 10\,keV-40\,MeV$ band, inconsistent with classical GRBs. 

    \item The time-resolved and time-integrated spectra are inconsistent with the Yonetoku and Amati spectral–energy relations for long GRBs.

    \item The spectra of episodes D and B, separated by $\sim$45\,minutes, are consistent in spectral shape and normalization.
    
    \item No significant high-energy cutoff is found for time-averaged spectra within the $\rm 10\,keV-40\,MeV$ range.
    
    \item The \textit{Fermi}/LAT ($E>100\rm\,MeV$) upper limits during the B and E episodes suggest that the spectrum softens between $10-100\rm\,MeV$. The combined D+B spectrum above 200\,keV further supports the presence of a high-energy softening.
    
    \item The constraints on the bulk Lorentz factor, variability, spectral shape and energetics make DBE consistent with the properties of four previously reported jetted TDEs.
    
\end{itemize}
If DBE originated from a jetted
TDE, it would represent the first such event with detected MeV emission. We argue that the observed spectra are most likely produced by synchrotron radiation from sub-TeV electrons.

\begin{acknowledgements}
We thank M.E. Ravasio, P. Blasi, P. Jonker and A. Levan for the fruitful discussions. We thank Fiona Harrison, PI of \textit{NuSTAR}, and the \textit{NuSTAR} SOC team for scheduling the requested DDT observation. F.A. acknowledges support from the Sichuan Science and Technology Department (under grant number 2024JDHJ0001). This research has made use of data obtained through the High Energy Astrophysics Science Archive Research Center Online Service, provided by the NASA/Goddard Space Flight Center, and specifically, this work made use of public Fermi-GBM and Fermi-LAT data. This work made use of data supplied by the UK Swift Science Data Centre at the University of Leicester. This research is based on observations obtained with \textit{NuSTAR}, a  project led by the California Institute of Technology, managed by the Jet Propulsion Laboratory, and funded by the National Aeronautics and Space Administration. Data analysis was performed using the \textit{NuSTAR} Data Analysis Software (NuSTARDAS), jointly developed by the ASI Science Data Center (SSDC, Italy) and the California Institute of Technology (USA).
\end{acknowledgements}

\bibliographystyle{aa}
\bibliography{bibliography} 

\begin{appendix}

\section{Fermi GBM data analysis}

\subsection{Orbital background estimate}\label{OSV}

To accurately estimate the background for the three observed $\gamma$-ray transients, we employed a modified version of the OSV tool \citep{De_Santis_2024, fitzpatrick_osv}. This method leverages the periodic orbital characteristics of the \textit{Fermi} satellite. \textit{Fermi} operates in 90-minute orbits and performs a rocking maneuver every two orbits to ensure full-sky coverage \citep{GBM}. This pattern results in the satellite returning to the exact same celestial position every 15 orbits, and to the same position with identical pointing every 30 orbits. The OSV tool exploits this by estimating the background for long or faint MeV sources as the mean event rate observed by the Gamma-ray Burst Monitor (GBM) 15 or 30 orbits before and after the time of interest.

For the binned data, the background estimate $R_{\rm bck}(t_{\rm i}, E_{\rm i})$ for an observed rate $R(t_{\rm i}, E_{\rm i})$ in a specific time and energy bin `i' is given by:
\begin{equation}
    R_{\rm bck}(t_{\rm i}, E_{\rm i}) = \frac{R(t_{\rm i} + 24\,{\rm h}, E_{\rm i}) + R(t_{\rm i}-24\,{\rm h}, E_{\rm i})}{2}\,.
\end{equation}
This formula applies to the 15-orbit scenario. For the 30-orbit scenario, the time difference becomes larger. Furthermore, averaging over 14 and 16 orbits also provides a robust background estimate \citep{fitzpatrick_osv}. This is attributed to the near-identical detector position and pointing angles at these orbital intervals.

It is crucial to note that the effectiveness of the 15-orbit method for background estimation, particularly above 40\,keV, is enhanced by the energy cutoff. Many potential point sources, which could introduce contamination, contribute predominantly below 100\,keV \citep{phys_model_gbm}. Consequently, for the majority of our energy bins, the position of the detectors is the dominant factor influencing the background, rather than subtle pointing variations.

\begin{figure*}[htbp]
    \centering
    \begin{subfigure}[b]{0.32\textwidth}
        \includegraphics[width=\textwidth]{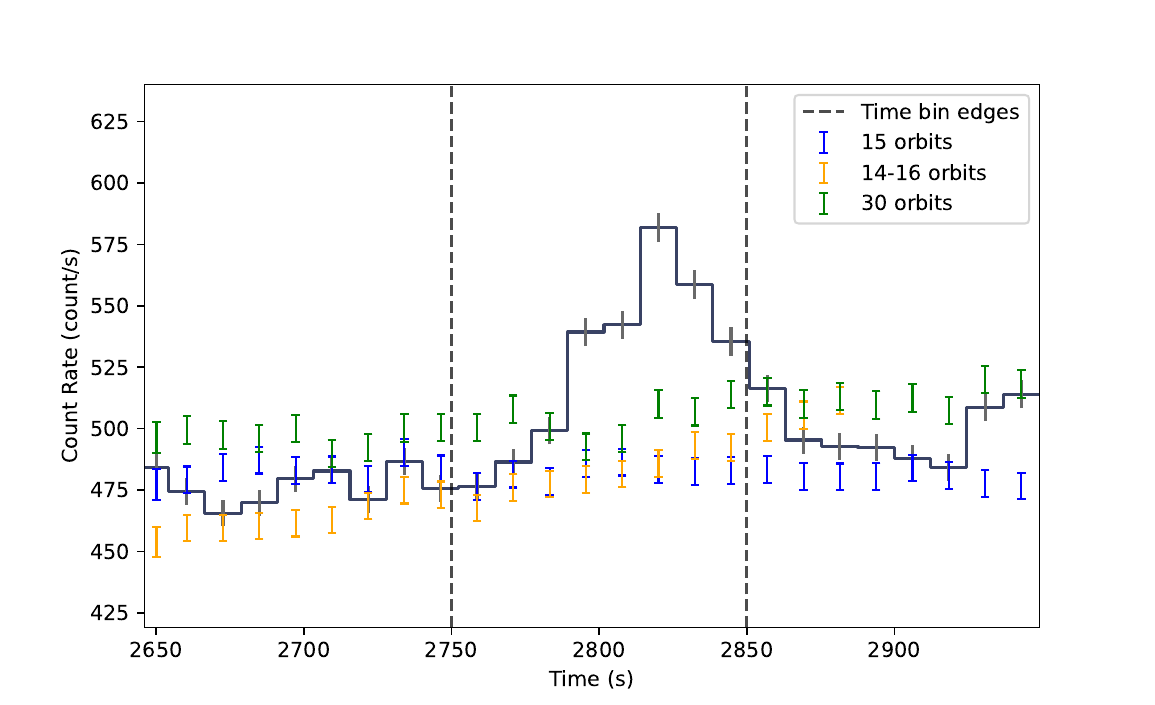}
        \caption{Detector nb, GRB B}
        \label{fig:d1e1}
    \end{subfigure}\hfill
    \begin{subfigure}[b]{0.32\textwidth}
        \includegraphics[width=\textwidth]{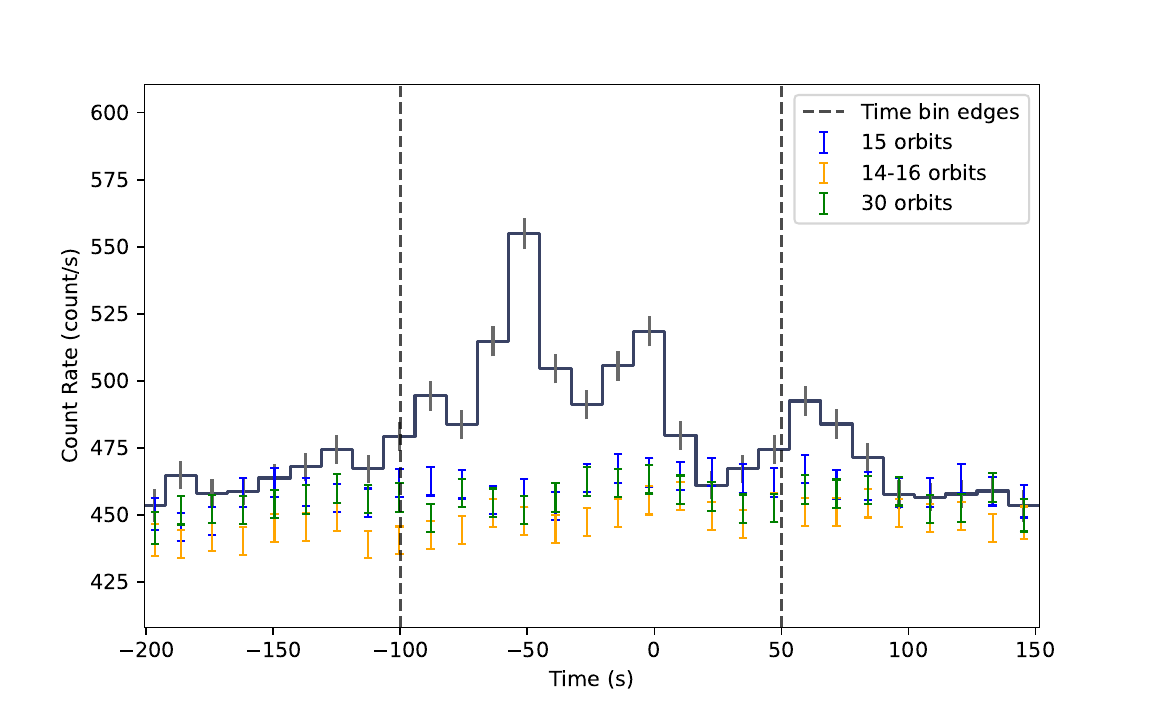}
        \caption{Detector nb, GRB D}
        \label{fig:d1e2}
    \end{subfigure}\hfill
    \begin{subfigure}[b]{0.32\textwidth}
        \includegraphics[width=\textwidth]{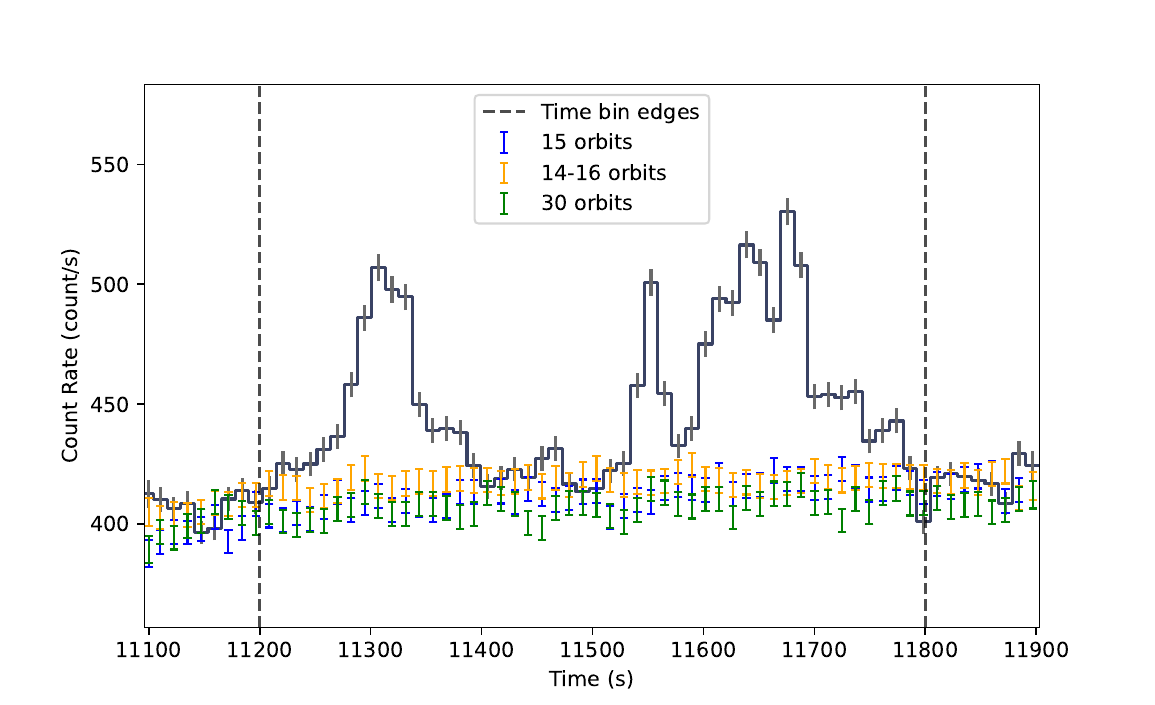}
        \caption{Detector nb, GRB E}
        \label{fig:d1e3}
    \end{subfigure}

    \vspace{-0.35em} 

    \begin{subfigure}[b]{0.32\textwidth}
        \includegraphics[width=\textwidth]{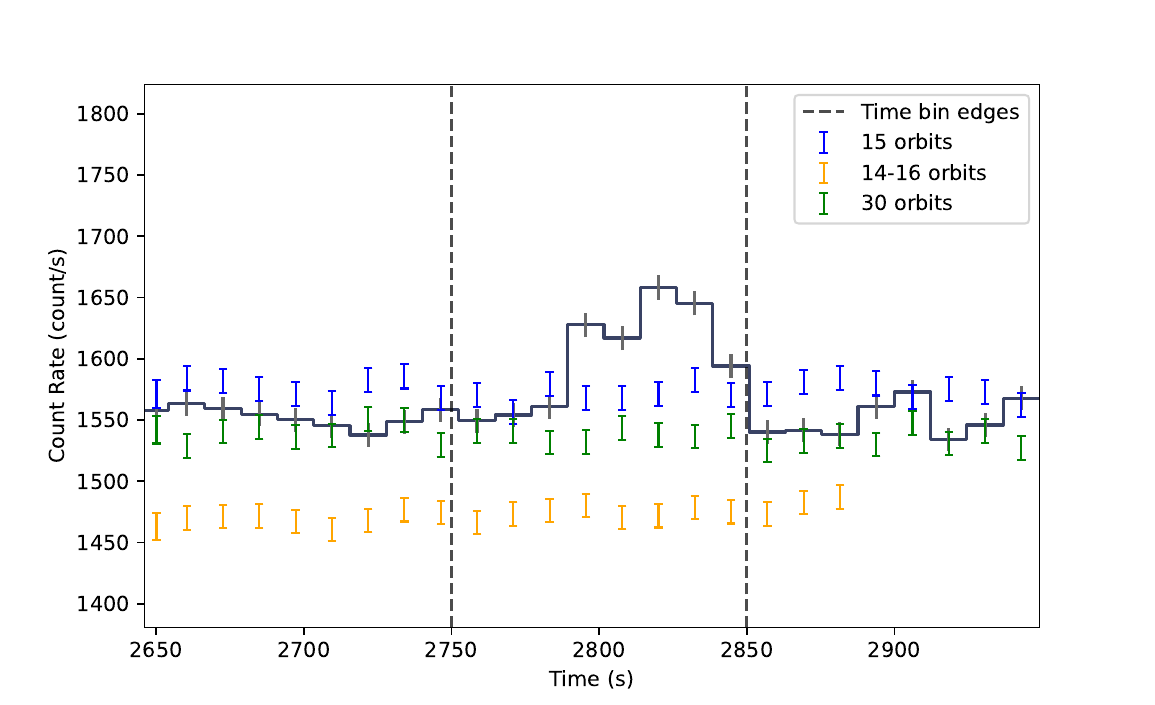}
        \caption{Detector b1, GRB B}
        \label{fig:d2e1}
    \end{subfigure}\hfill
    \begin{subfigure}[b]{0.32\textwidth}
        \includegraphics[width=\textwidth]{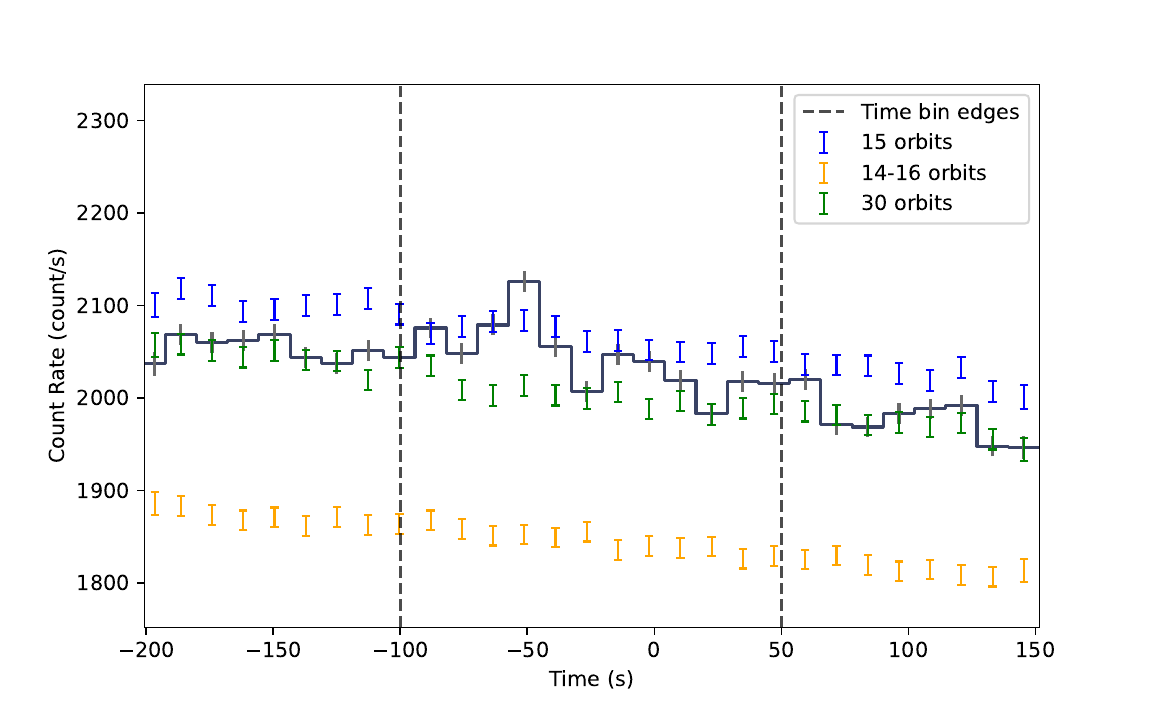}
        \caption{Detector b1, GRB D}
        \label{fig:d2e2}
    \end{subfigure}\hfill
    \begin{subfigure}[b]{0.32\textwidth}
        \includegraphics[width=\textwidth]{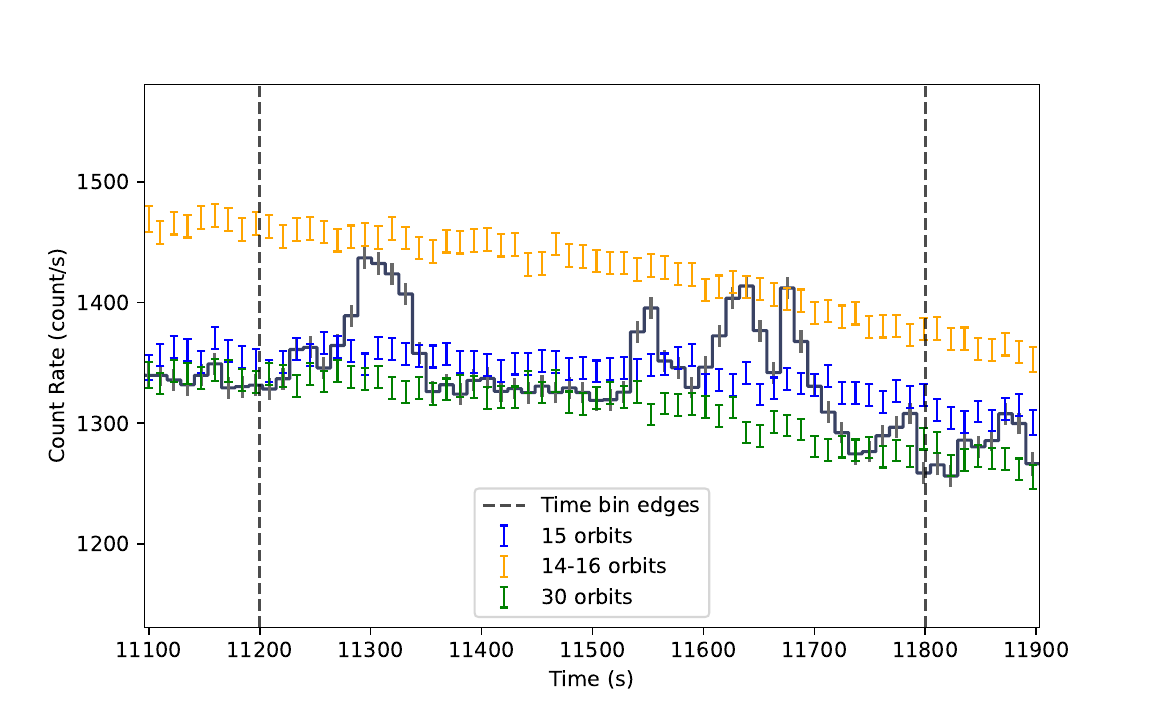}
        \caption{Detector b1, GRB E}
        \label{fig:d2e3}
    \end{subfigure}

    \caption{Light curves for two detectors (Detector nb: top row; Detector b1: bottom row) across three different events (Event 1: left column; Event 2: middle column; Event 3: right column). The estimated background is plotted using three different orbits. For the `nb' data, the energy range is cut between 40 keV and 900 keV, while for `b1' data, the cut is between 200 keV and 40 MeV. Each panel shows the light curve integrated over energy for a specific detector and event combination.}
    \label{fig:combined_lightcurves}
\end{figure*}

Fig.\,\ref{fig:combined_lightcurves} illustrates the background estimates for the primary time intervals for detectors `nb' and `b1'. A qualitative validation of the background estimation can be done by comparing the predicted and measured rates in regions immediately adjacent to the signal. A close match indicates a valid background estimate. 

Analysis of the data reveals that the background estimated using the 30-orbit method generally aligns well with the measured rates. The sole exception is detector `nb' during GRB B, where the 30-orbit estimate clearly overestimates the background. In this specific case, the 15-orbit estimate provides a much more accurate representation of the pre- and post-burst background.

\subsection{Spectral analysis}\label{GBMspectra}

We first verified that during the temporal bins of the D, B, and E events, only one sodium iodide detector (nb) and one bismuth germanate detector (b1) satisfy the source-to-detector angle cuts. We downloaded the daily data for July 2, 2025, from the \textit{Fermi} GBM Daily Data online repository\footnote{\url{https://heasarc.gsfc.nasa.gov/W3Browse/fermi/fermigdays.html}}. To analyze the time bins of interest, we generated response matrices using the GBM Response Generator\footnote{\url{https://fermi.gsfc.nasa.gov/ssc/data/analysis/gbm/DOCUMENTATION.html}}, based on the position provided by \textit{Swift}/XRT. We identify the D, B, and E outburst times as $-100$--$50$ s, $2750$--$2850$ s, and $11200$--$11800$ s, respectively, since the trigger time of GRB 250702D (2 July 13:09:02.03 UT). We extracted the spectra from GBM daily data using the publicly available \textit{Fermi} GBM Data Tools \citep{GbmDataTools}. 

We tested two methods to estimate the background spectrum during the D, B, and E intervals (see below). The first method extrapolates a polynomial function fitted to pre- and post-burst intervals, allowing background estimation over the full \textit{Fermi}/GBM energy range ($\rm 10\ keV-40\ MeV$). The second method uses the average orbital background, which lacks coverage below 40\,keV. The results of the spectral modeling are presented in Table~\ref{tab:spectral_fits}. We note that all spectral fits presented in this work are performed using \texttt{XSPEC} \citep{Arnaud1996}, and assuming Cash statistics ($C$-stat).

\subsubsection{Spectra of events D and B}

The duration of the D event is short ($\sim$150 s), so we estimate the background using both the standard polynomial fit to off-source intervals ($-300$ to $-200$\,s and $150$ to $250$\,s), and the orbital background method (above 40\,keV). The model parameters from both methods are consistent within uncertainties. We report the photon index and flux ($10{\rm\, keV}$–\,$40 {\rm\,  MeV}$) for the power-law model, and the cutoff energy $E_{\rm cut}$ for the cutoff power-law model. We do not observe a significant statistical improvement from introducing a high-energy cutoff ($\rm \Delta\ stat = -4$, for one additional free parameter).

Similarly to D, event B also has a short duration ($\sim$150 s), so both background estimation methods are applied, yielding comparable results. For wider energy coverage, we use the standard off-source intervals ($2350$–$2550$\,s and $3000$–$3055$\,s). The cutoff power-law model gives only a marginal improvement in fit quality ($\rm \Delta\ stat = -14$, for one additional free parameter).

Given that the background at the highest energies ($>$200 keV) is well described by the orbital method, and that the detector response remains stable between events D and B, we estimate the cumulative D+B spectrum ($>$200 keV, detector b1) for improved spectral characterization. The corresponding power-law fit is included in Table~\ref{tab:spectral_fits}.

\subsubsection{Spectrum of event E}

The E outburst lasts at least 600 s. Therefore, we adopt the orbital background method (30 orbits). We model the spectrum using both power-law and cutoff power-law models. The latter does not significantly improve the spectral characterization in the $\rm 40\ keV-40\ MeV$ range ($\Delta$stat = 1).

\begin{table*}[ht]
\centering
\begin{tabular}{l c | c c c | c c}
\hline
\multirow{2}{*}{Name} &\multirow{2}{*}{Time-bin [s]} & \multicolumn{3}{c|}{Power-law model} & \multicolumn{2}{c}{Cutoff Power-law} \\
\cline{3-7}
 &  & Photon index & Flux [$10^{-7}{\rm\; erg}{\rm\; cm}^{-2}{\rm\; s}^{-1}$] & stat(dof) & $E_{\rm cut}$ [MeV] & stat(dof) \\
\hline
D     & $(-100,\,50)$      & $1.48 \pm 0.06$        & $2.7^{+0.6}_{-0.4}$     & 208(235) & $>4$              & 204(234) \\[1ex]
B     & $(2750,\,2850)$    & $1.47 \pm 0.05$        & $2.7^{+0.4}_{-0.3}$     & 253(235) & $7^{+13}_{-6}$   & 239(234) \\[1ex]
E & $(11200,\,11800)$  & $1.64 \pm 0.04$                    & $0.56^{+0.04}_{-0.05}$  & 289(215) & $>14$             & 288(214) \\[1ex]
D+B &   & $1.77^{+0.17}_{-0.12}$                    & $1.61^{+0.08}_{-0.50}$  & 125(119) &              &  \\[1ex]
\hline
\end{tabular}
\caption{Spectral fit results for selected time bins using power-law and cutoff power-law models. Data between 10\,keV and 40\,MeV and the standard off-source background estimate is applied to D and B. For the spectrum E, an average orbital method is applied and thus spectrum at $\rm 40\ keV - 40\ MeV$ is extracted. For the cumulative D+B spectrum, we have used the average orbital background method and the spectrum at $ \rm 200\ keV - 40\ MeV $ is extracted. The reported flux is evaluated at the covered energy band.}
\label{tab:spectral_fits}
\end{table*}

\section{X-rays}\label{Xrays}
\subsection{Swift/XRT data analysis}

To extract the spectra, we downloaded the XRT event files from the Swift-XRT archive\footnote{\url{http://www.swift.ac.uk/archive/}}.  We extracted source and background spectra in each time bin using the \texttt{xselect} tool. For each time bin, the ancillary response has been generated by \texttt{xrtmkarf}. We then model 9 spectra with an absorbed power-law model with two neutral hydrogen absorbers: one fixed to Galactic the equivalent hydrogen column density at $N_{\rm H} = 0.207 \times 10^{22}\,\rm cm^{-2}$ and the other at the host galaxy. We have adopted 4 different redshifts for the host galaxy: 0, 0.5, 1 and 2. The equivalent hydrogen column density in the host galaxy is best fitted by $N_{\rm H}^{z=0} = (0.8 \pm 0.2)  \times 10^{22}\,\rm cm^{-2}$, $N_{\rm H}^{z=0.5} = (2.2 \pm 0.6)  \times 10^{22}\,\rm cm^{-2}$ , $N_{\rm H}^{z=1} = (4.6 \pm 1.0)  \times 10^{22}\,\rm cm^{-2}$, and $N_{\rm H}^{z=2} = (13.0 \pm 3.0)  \times 10^{22}\,\rm cm^{-2}$. We have verified that the inferred unabsorbed flux is not affected by the choice of the redshift. And so, we report the unabsorbed $0.3-10$\,keV flux ($z=0$) and the photon index in the Table\,\ref{table:XRT}.  

\begin{table}
\centering
    \begin{tabular}{cccc}
    
        \hline
        $T_\mathrm{start}$ & $T_\mathrm{stop}$ & Flux & Photon Index \\
        (days) & (days) & ($10^{-11}$ erg\,cm$^{-2}$\,s$^{-1}$) & \\
        \hline
        
        0.54 & 0.55 & 12.35\(_{-1.30}^{+1.50}\) & 1.61\(_{-0.22}^{+0.23}\) \\ \\[-1.5ex]
        0.58 & 0.59 & 3.53\(_{-0.62}^{+0.84}\) & 1.90\(_{-0.34}^{+0.36}\) \\ \\[-1.5ex]
        0.73 & 0.74 & 7.59\(_{-0.91}^{+1.15}\) & 1.78\(_{-0.23}^{+0.24}\) \\ \\[-1.5ex]
        0.795 & 0.798 & 3.50\(_{-0.71}^{+0.85}\) & 1.37\(\pm 0.40\) \\ \\[-1.5ex]
        1.30 & 1.31 & 1.64\(_{-0.36}^{+0.45}\) & 1.40\(_{-0.47}^{+0.47}\) \\ \\[-1.5ex]
        1.37 & 1.39 & 1.33\(_{-0.20}^{+0.28}\) & 1.90\(_{-0.31}^{+0.33}\) \\ \\[-1.5ex]
        1.44 & 1.45 & 1.33\(_{-0.25}^{+0.32}\) & 1.83\(_{-0.38}^{+0.40}\) \\ \\[-1.5ex]
        2.22 & 2.36 & 0.39\(_{-0.08}^{+0.11}\) & 1.90\(_{-0.38}^{+0.40}\) \\ \\[-1.5ex]
        5.33 & 5.41 & 0.10\(_{-0.04}^{+0.17}\) & 2.24\(_{-1.22}^{+1.25}\) \\ \\[-1.5ex]
        \hline
    \end{tabular}
    
\caption{The result of the time-resolved spectral analysis of \textit{Swift}/XRT data.}
\label{table:XRT}
\end{table}

\subsection{NuSTAR data analysis}

\begin{figure}[h]
    \centering
    \includegraphics[width=0.48\linewidth]{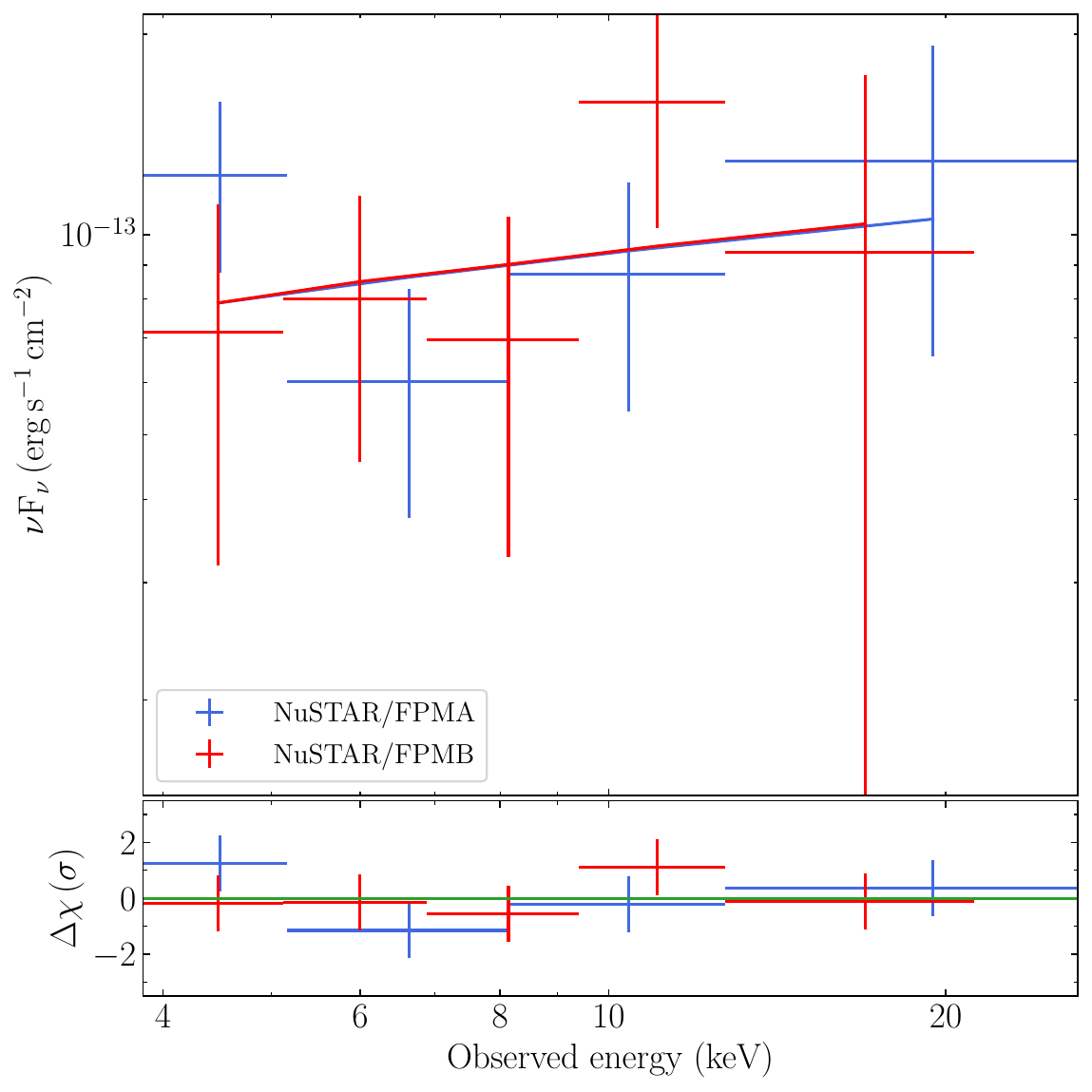}
    \includegraphics[width=0.48\linewidth]{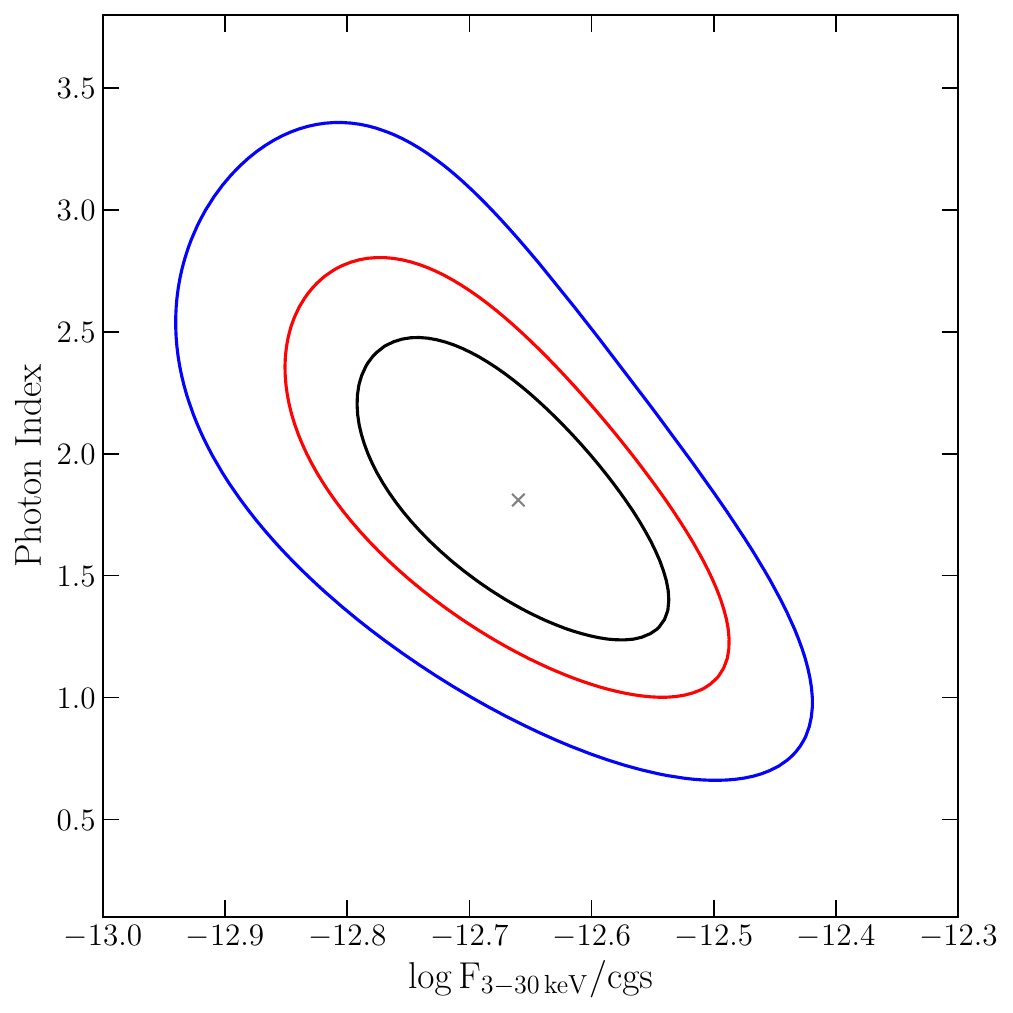}
    \caption{Left: \textit{NuSTAR} FPMA and FPMB (blue and red, respectively) spectra fitted with a power-law model. Right: The confidence contours (68\%, 95\%, and 99\%) of the photon index versus the $3-30$\,keV flux.}
    \label{fig:nustar}
\end{figure}

A \textit{NuSTAR} \citep{Harrison13} DDT observation of DBE took place on 2025 July 12 (ObsID 91101324002; PI: E. Kammoun) for a total exposure of 39\,ks. The data were reduced using the standard pipeline in the NuSTAR Data Analysis Software (\textsc{NuSTARDAS v2.1.4}), and using \texttt{CalDB20250521}. We cleaned the unfiltered event files with the standard depth correction. We reprocessed the data using the \texttt{saamode = strict}, \texttt{saacalc=3}, and \texttt{tentacle = yes} criteria for a more conservative treatment of the high background levels near the South Atlantic Anomaly. Due to the high background activity during the observation, this filtering scheme reduced our net exposure time to 31.7\,ks and 32.8\,ks, for the two focal plane modules (FPMA and FPMB), respectively. We extracted the source spectra from a circular region of radius 40\arcsec\ centered at the coordinates of the source, using the HEASOFT task \texttt{nuproducts}. Background spectra were extracted from a source-free region of radius 80\arcsec. We analyze the spectra from FPMA and FPMB jointly, without combining them. The \textit{NuSTAR} spectra were binned using the ftool \texttt{ftgrouppha} with \texttt{grouptype=optmin} and \texttt{groupscale = 25}. This sets up a binning using optimal binning as described by \cite{kaastra16}, but with the additional requirement of a minimum number of 25 counts per bin.

We modeled the \textit{NuSTAR} spectra in the $3-30$\,keV range using \texttt{XSPEC}. We assumed an absorbed power law, by fixing the equivalent hydrogen column density at $N_{\rm H} = 10^{22}\,\rm cm^{-2}$ (assumed at $z=0$, consistent with the analysis of \textit{Swift}/XRT). We let the photon index and the normalization free to vary. This model results in a statistically good fit ($C/\rm dof = 5/7$)with a photon index of $1.8_{-0.6}^{+0.7}$ and $3-30$\,keV flux $\log F_{3-30} = -12.66 \pm 0.14$. Extrapolating the best fit to the \textit{Swift}/XRT soft band results in an unabsorbed flux in the $0.3-10$\,keV range of $\log F_{0.3-10} = -12.6_{-0.3}^{+0.5}$. The left panel of Fig.\,\ref{fig:nustar} shows the \textit{NuSTAR} spectra with the best-fit model. The right panel of the same figure shows the photon index versus flux contours.

\section{Fermi/LAT}\label{LAT}
We performed unbinned likelihood analysis of \textit{Fermi}/LAT data for the $\rm{\gamma}$-ray burst, starting from the 5\,ks before the \textit{Fermi}/GBM trigger time of GRB~250702D (t$_{0}$=773154547 s MET) and expanding up to t$_{0}+$15\,ks seconds, in the energy range of 0.1–1 GeV, using the \texttt{GTBURST} \footnote{\url{https://fermi.gsfc.nasa.gov/ssc/data/analysis/scitools/gtburst.html}} software from Fermi Science tool. We selected the region of interest covering $12^{\circ}$ around the source location R.A. = 284.69$^\circ$, Dec=-7.87$^\circ$ (J2000) consistent with the \textit{Swift}/XRT and EP observations \citep{EP,Swift}. A standard zenith angle (Zd) cut of 100$^{\circ}$ is applied to remove any contamination of photons from the Earth-limb. We used the \texttt{P8R3$\_$TRANSIENT020} event class, which is suitable for the analysis of transient sources, and the corresponding instrument response functions. The isotropic particle background (\texttt{`isotr template'} in \texttt{GTBURST}), galactic and extragalactic high-energy components of the Fermi Fourth Catalog (4FGL), with fixed normalization (\texttt{`template (fixed norm)'}) has been used.  No significant emission during the transients was detected. We estimated a 3$\sigma$ (99.7\%) upper-limit of the energy-flux while fitting a power-law (\texttt{`powerlaw2'} spectral model using \texttt{GTBURST}). 

\begin{figure}[h]
    \centering
    \includegraphics[width=0.90\linewidth]{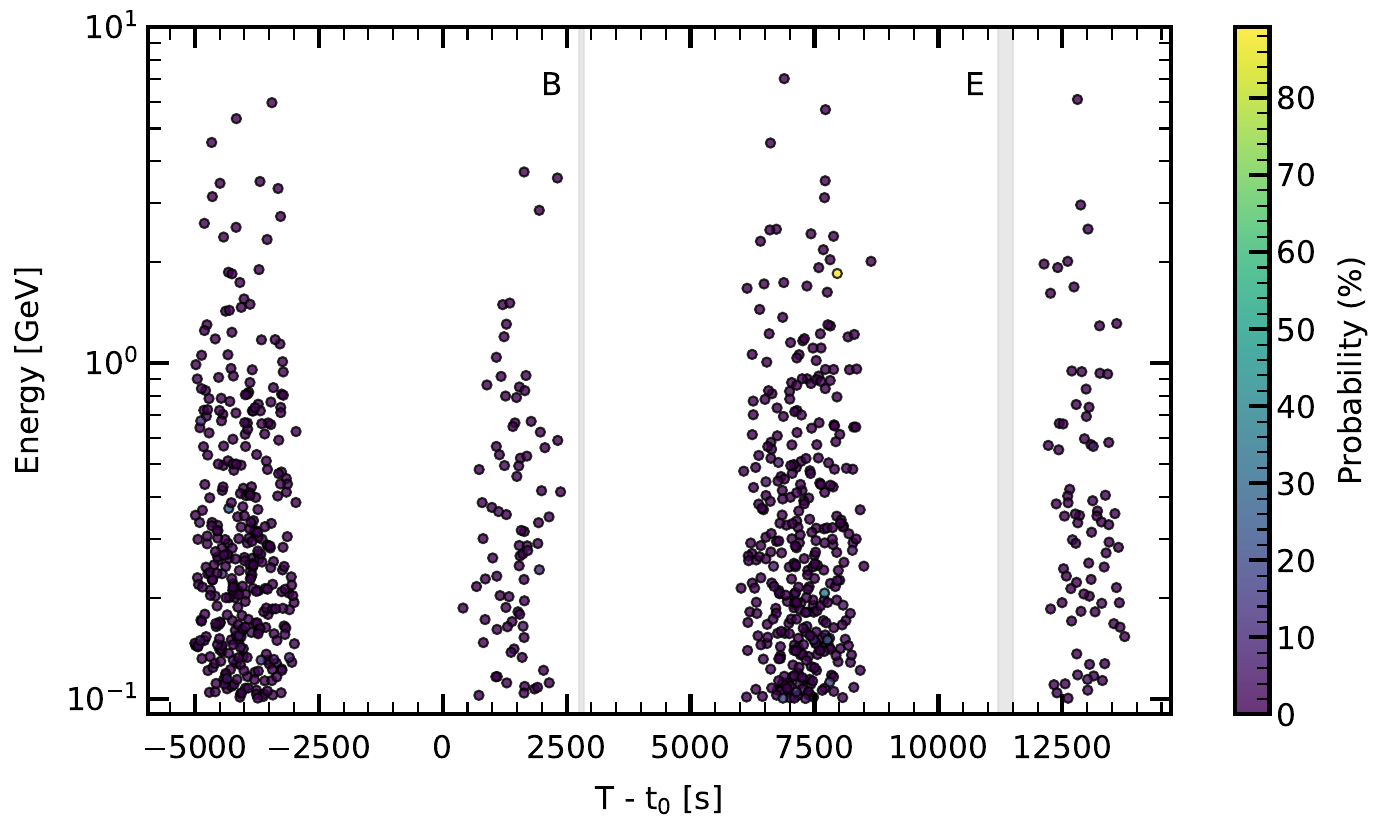}
    \caption{GeV photon detected by LAT from the location of GRB~250702D/B/E and probability of association with the transient. The figure shows the time-windows in which high-energy photons were detected by LAT from the transient location. The color bar represents the probability of association of photons to the transient detected during $t_{0} - 5$\,ks to $t_{0} + 15$\,ks. The gray shaded region represents GRB 250702B/E. Note that LAT observed event E for approximately half of its duration. Thus, the time interval marked by the shaded region corresponds to time when LAT covered the transient.}
    \label{fig:LAT_photon}
\end{figure}

In addition, we computed the probability of association of photons with the burst in the given time interval using the \textit{gtsrcprob} module in an energy range of 0.1-100\,GeV. Figure~\ref{fig:LAT_photon} shows the photons detected by LAT during this period and the probability of photons associated with the transient. We detected only a single photon with an energy of 1.85\,GeV with a probability greater than 88\% of being associated with the source.

Due to the unavailability of good quality data during GRB 250702D (not satisfying Zd cut; Zd$>$100$^{\circ}$), and no high-energy photons registered by LAT (even if the observation condition of Zd$<$100$^{\circ}$ was satisfied) during GRB 250702B and GRB 250702E (see Tab.~\ref{tab:spectral_fits} and Fig.\ref{fig:LAT_photon}), we selected a time-range starting from $-4800$\,s before the trigger to derive the flux upper-limits for two different durations of 100\,s and 300\,s (see Tab. \ref{tab:gev_analysis}) in the energy range 0.1-1 GeV. 

\begin{table*}[h]
    \centering
    \begin{tabular}{|c|c|c|c|}
        \hline
        {Duration} {[s]} & Analysis period (T-t$_{0})$ & $\Gamma=-2$ [erg cm$^{-2}$ s$^{-1}$] & $\Gamma=-2.5$ [erg cm$^{-2}$ s$^{-1}$] \\ 
        \hline
        100 & {-4800, -4700} & 6.8$\times10^{-9}$ & 6.7$\times10^{-9}$ \\ 
        \hline
        300 & {-4800, -4500} & 3$ \times 10^{-9}$ &  3.1$\times10^{-9}$\\ 
        \hline
        
    \end{tabular}
    \caption{Upper limits in the $0.1-1$\,GeV range at the  3$\sigma$ level. Given that DBE was not covered with LAT during the burst periods, we estimated these upper limits one orbit before at the same position of the sky for two different exposures of 100\,s and 300\,s.}
    \label{tab:gev_analysis}
\end{table*}

\section{Physical constraints}\label{physics}

We briefly discuss the possibility that the observed $\gamma$-ray emission is produced through synchrotron and inverse Compton (IC) radiation by a population of cooled electrons. The properties of the emission region can be constrained by requiring that (i) the radiative cooling time is shorter than the dynamical time, and (ii) the jet has a high radiative efficiency (i.e., a large fraction of the jet energy is converted into MeV $\gamma$-rays).

Producing the observed emission through synchrotron radiation requires electron Lorentz factors $\gamma_e \sim 10^7 \Gamma_1^{-1/2} (\epsilon/\epsilon_{\max})^{1/2} B^{\prime -1/2}$, where $B^{\prime}$ is the magnetic field in the comoving frame of the emission region and $\Gamma_1=\Gamma/10$. The synchrotron cooling time scale, $t^\prime_{\rm sync}\sim 6\pi m_e c/\gamma_e B^{\prime 2}\sigma_{\rm T}\sim 80\, \Gamma_1^{1/2} (\epsilon/\epsilon_{\max})^{-1/2} B^{\prime -3/2} {\rm\, s}$ should be shorter than the dynamical time in the comoving frame, $t^{\prime}_{\rm dyn}\sim R/c\Gamma \sim \Gamma \delta t_v\sim 10^3 \, \Gamma_1 \delta  t_{2}{\rm\, s}$, where $R\sim \Gamma^2 c\delta t_v$ is the distance from the compact object where the $\gamma$-ray emission is produced and $\delta t_{2}=\delta t_v/100{\rm\, s}$. For $\epsilon= \epsilon_0 \sim 2.6\times 10^{-4}\epsilon_{\max}$, the condition $t^\prime_{\rm sync}<t^\prime_{\rm dyn}$ implies $B^{\prime} > 3 \, \Gamma_1^{-1/3} \delta t_{2}^{-2/3} {\rm\, G}$.
The jet Poynting flux, $L_{\rm B}\sim c\Gamma^2 B^{\prime 2} R^2$, should be smaller than the isotropic-equivalent outburst luminosity, $L_{\rm iso}$. The condition $L_{\rm B}<L_{\rm iso}$ implies $B^{\prime}< 2\times 10^3 L_{48}^{1/2} \Gamma_1^{-3} \delta t_2^{-1} {\rm\, G}$, where $L_{48}\sim L_{\rm iso}/10^{48}{\rm\, erg\, s}^{-1}$. Therefore, a population of sub-TeV electrons ($5\times 10^3 \, \Gamma_1^{-1/2} B_3^{\prime -1/2} < \gamma_e < 3 \times 10^5\, \Gamma_1^{-1/2} B_3^{\prime -1/2}$, where $B_3^{\prime}=B^{\prime}/10^3{\rm\, G}$) can account for the observed spectra of DBE. The maximum energy of electrons can be roughly estimated by requiring the acceleration time $t_{acc}^{\prime}\approx \eta r_{L}/c$ to be compatible with $t_{syn}^{\prime}$, where $r_{L}$ is the Larmor radius of an electron. As a result, $\gamma_{max} \sim \left(\frac{6\pi q_e}{\eta B^{\prime} \sigma_T}\right)^{1/2}$, and for $\eta \sim 1$ we obtain $\gamma_{max} \sim 4 \times 10^{6} B_{3}^{\prime -1/2}$, which is consistent with our earlier requirement based on radiative efficiency.

Producing the observed MeV radiation via inverse Compton (IC) scattering is complicated. To maintain a high radiative efficiency in MeV $\gamma$-rays, the lowest-energy electrons should scatter the lowest-energy observed photons ($\epsilon=\epsilon_0$, which corresponds to an energy $\sim$10 keV) in the Klein-Nishina regime. Then, the Lorentz factor of the lowest-energy electrons should be $\gamma_e > 5 \times 10^{2} \,\Gamma_{1}$. The Lorentz factor of the highest-energy electrons should be $\gamma_e > 3 \times 10^{4} \, \Gamma_{1}$ because $\epsilon_{\max}/\epsilon_0 \sim 4 \times 10^{3}$. Therefore, seed photons should have energies $< 2 \times 10^{-2} \, \Gamma_{1}^{-2}$\,eV. In the external IC scenario, it is difficult to produce far-infrared photons from the accretion disk. In the synchrotron self-Compton scenario, one would need an unreasonably low magnetic field, $B^{\prime} < 0.1 \ \Gamma_{1}^{-5}\rm\, mG$.

\section{Isotropic-equivalent energy}

\begin{figure}[ht]
\centering
\includegraphics[width=0.70\linewidth]{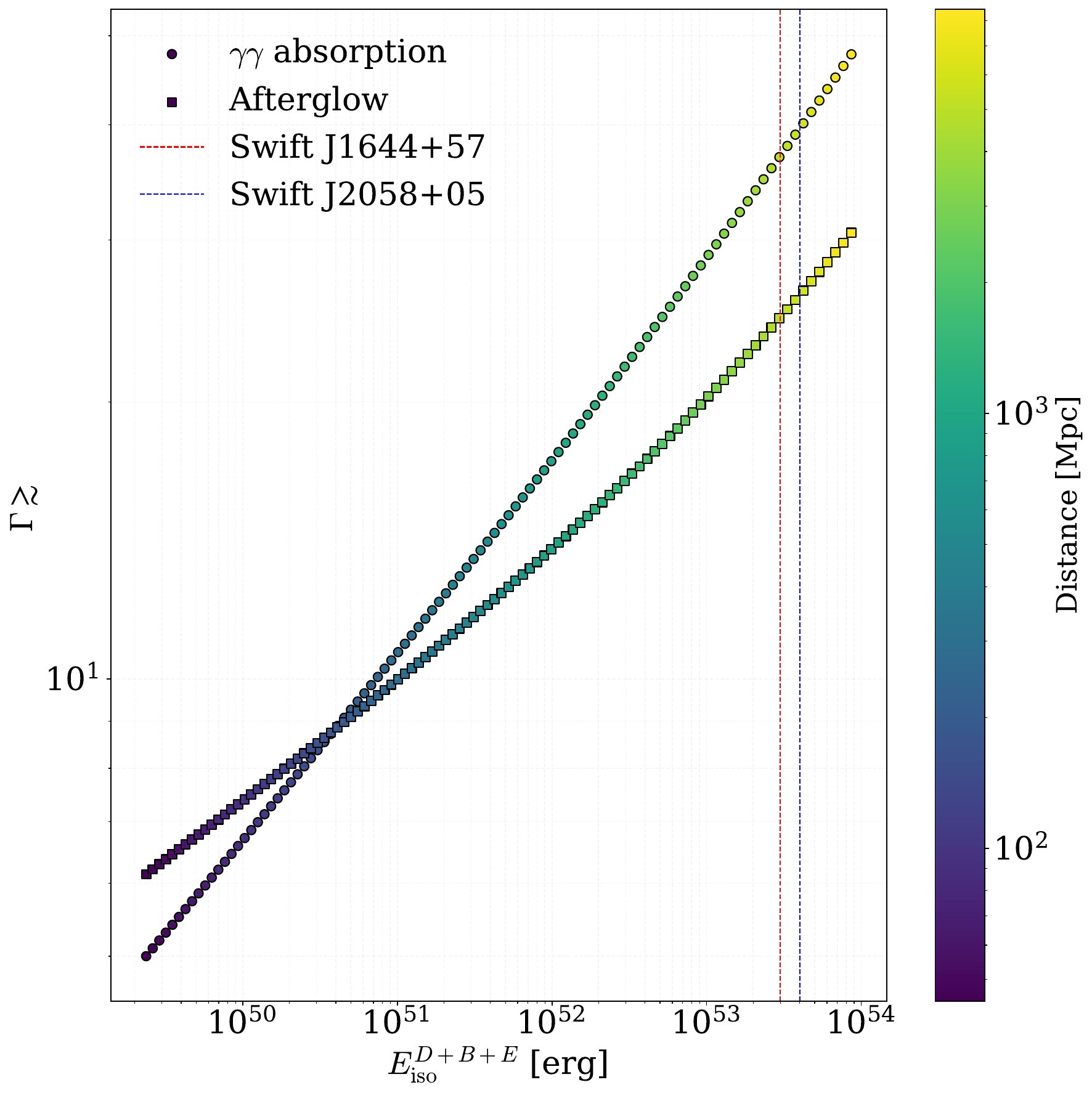}
\caption{Lower limit on the bulk Lorentz factor as a function of the isotropic-equivalent energy released during all three episodes. The lower limits are determined using two methods: the afterglow deceleration time (squares) and $\gamma$--$\gamma$ absorption (circles).} The color bar indicates the corresponding luminosity distance to DBE. For comparison, we show total isotropic-equivalent energy measured for Swift J2058+05 \citep{Cenko2012,Pasham2015} and Swift\,J1644+57 \citep{Burrows2011} as dashed blue and red vertical lines.
    \label{fig:Gamma2}
\end{figure}

Fig.~\ref{fig:Gamma2} shows the lower limit on the bulk Lorentz factor of the jet as a function of the isotropic-equivalent energy released during the three emission episodes. These lower limits are inferred from two independent methods: $\gamma$--$\gamma$ absorption (circles) and constraints from the deceleration time of the X-ray afterglow emission. The latter is estimated using the standard deceleration time condition in a uniform circumburst medium, $\Gamma > (3 E_{0}/2^{5}\pi n_{0} m_p c^{5} t_{p,z}^{3})^{1/8}$ \citep{Piran1999}, where $n_0$ is the number density of the circumburst medium, $E_0$ is the isotropic-equivalent kinetic energy of the jet, and $t_{p,z}$ is the afterglow peak time in the source rest frame. For the limits shown here, we assume a radiative efficiency of 10\% for the prompt MeV emission, $n_0 = 1~\mathrm{cm^{-3}}$, and $t_p < 0.5$ days.

\section{Comparison with other relativistic TDE candidates}\label{comparison}

In Table~\ref{tab:comparison}, we summarize the properties of four previously reported relativistic TDE candidates, along with the DBE event. The characteristics of DBE closely resemble those of these events in terms of X-ray temporal decline, spectral index, peak luminosity, and isotropic-equivalent energy. All of the sources have been detected in hard X-rays (15–150 keV), with the exception of AT2020cmc. The typical variability timescale of DBE ($\sim 100$ s) is comparable to the shortest variability reported for Swift J1644+57 ($\sim 100$ s) and Swift J2058+05 ($\sim 500$ s). Estimates of the bulk Lorentz factor in TDE jets are often model dependent but generally fall within the range 2–90.

\begin{table*}[ht]
\centering
\renewcommand{\arraystretch}{1.6} 
\setlength{\tabcolsep}{4pt}       
\begin{tabular}{lcccccccc}
\hline
Source 
& $t^{-\alpha_X}$ 
& $\Gamma_X$ 
& $L_{X,\mathrm{iso}}$ & $E_{X,\mathrm{iso}}$ & $t_v$ & $\Gamma$ & $z$ \\
&  &  & (erg\,s$^{-1}$) & (erg) & (s) &  &  \\
\hline
Swift J1644+57 
& agree with $t^{-5/3}$[1] 
& $1.62 \pm 0.05$[$\star$,2] 
& $3 \times 10^{48}$[2,3] 
& $2 \times 10^{53}$[3] 
& \makecell{$\sim 100$[2] \\ $\sim 105$[3]} 
& $2 < \Gamma < 20$ 
& 0.35 \\
Swift J2058+05 
& $t^{-2.2}$[4] 
& \makecell{$1.61 \pm 0.12$[4] \\ $1.47 \pm 0.08$[5]} 
& $3 \times 10^{47}$[4] 
& \makecell{$\approx 10^{54}$[4] \\ $\approx 4 \times 10^{53}$[5]} 
& \makecell{$< 10^{4}$[4] \\ $\sim 500$[5]} 
& $> 2.1$[4] 
& 1.19 \\
Swift J1112$-$82 
& $t^{-1.1}$[$\star\star$,6] 
& $1.33 \pm 0.08$[6] 
& $\geq 10^{47}$[6] 
& -- 
& -- 
& -- 
& 0.89 \\
AT2020cmc 
& $t^{-2 \pm 0.1}$[7] 
& \makecell{$1.4 \pm 0.2$[7] \\ $1.45 \pm 0.06$[8]} 
& \makecell{$\approx 2.4 \times 10^{47}$[7] \\ $\ge 10^{48}$[8]} 
& -- 
& \makecell{$\sim 4 \times 10^{3}$[7] \\ $\le 10^{3}$[8]} 
& \makecell{$\approx 12$[7] \\ $\approx 86$[8] [$\star\star\star$]} 
& 1.19 \\
\hline
DBE 
& $t^{-1.9 \pm 0.1}$ 
& \makecell{$\Gamma_X \sim 1.4$--$1.6$ \\ $\Gamma_\gamma \sim 1.5$} 
& $\approx 4 \times 10^{47}\, d_{L,100\,\mathrm{Mpc}}^{2}$ 
& $\approx 10^{50}\, d_{L,100\,\mathrm{Mpc}}^{2}$ 
& $\le 100$ 
& $10 < \Gamma < 40$ 
& -- \\
\hline
\end{tabular}
\caption{Properties of relativistic TDE candidates. 
[1]~\cite{Burrows2011}, 
[2]~\cite{Levan2011}, 
[3]~\cite{Bloom2011}, 
[4]~\cite{Cenko2012}, 
[5]~\cite{Pasham2015}, 
[6]~\cite{Brown2015}, 
[7]~\cite{Andreoni2022}, 
[8]~\cite{Pasham2023}.  
[$\star$]~XRT+BAT joint spectrum (0.3--150 keV).  
[$\star\star$]~Followed by the steep decline $>$30 days.  
[$\star\star\star$]~In the synchrotron self-Compton model.}
\label{tab:comparison}
\end{table*}

\section{Comparison with spectral-energy correlations of GRBs}\label{correlations}

\begin{figure*}[htbp] 
    \centering 
    \begin{subfigure}[b]{0.400\textwidth} 
        \centering
        \includegraphics[width=\textwidth]{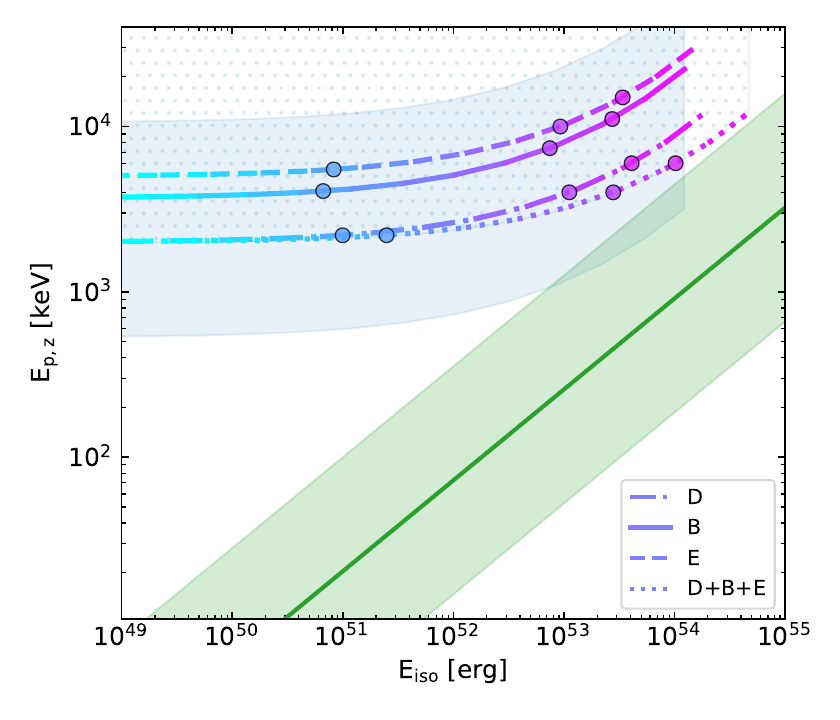} 
        \caption{
        }
        \label{fig:amati}
    \end{subfigure}
    \begin{subfigure}[b]{0.456\textwidth}
        \centering
        \includegraphics[width=\textwidth]{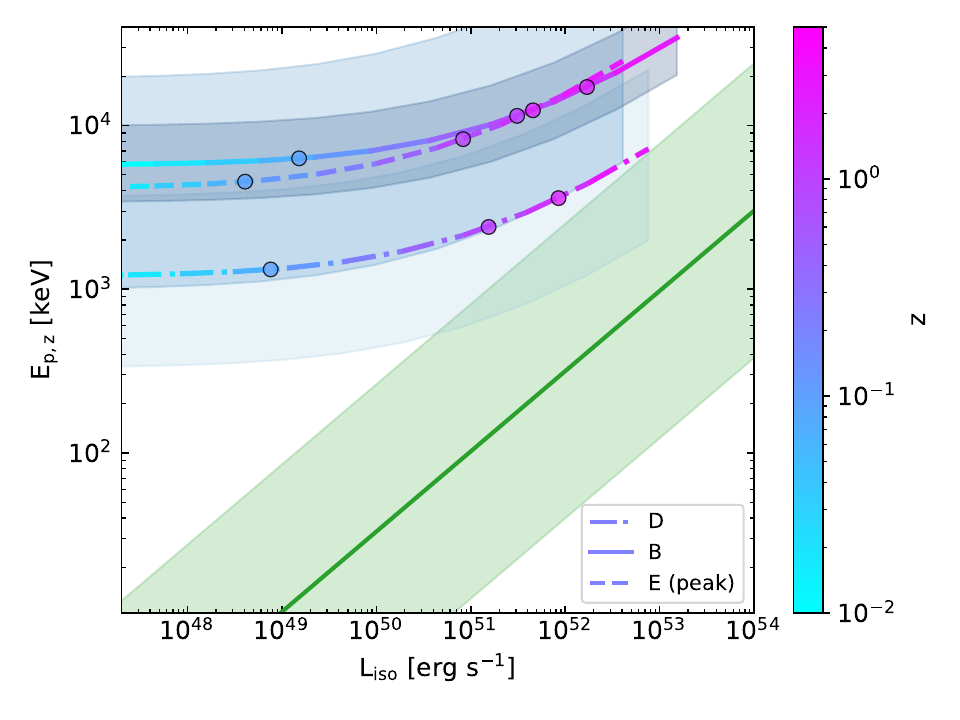} 
        \caption{
        }
        \label{fig:yonetoku}
    \end{subfigure}    
    \caption{Comparison of the intrinsic properties of DBE, computed for different redshifts, with the Amati relation (a) and the Yonetoku relation (b). The green line in each panel shows the corresponding relation, with the green-shaded area indicating its $3\sigma$ scatter. The colored lines display $E_{p,z}$ as a function of $E_{\mathrm{iso}}$ (left) and $L_{\mathrm{iso}}$ (right), computed for spectra D, B, and E (distinguished by line style) as a function of $z$, encoded by the color map. Circles on top of each curve denote reference redshift values, $z = 0.1, 1, 2$. The blue-shaded bands indicate the $1\sigma$ confidence intervals (for spectrum B in the left panel; for spectra D, B, and E in the right panel), while the blue dots denote regions corresponding to lower limits on the fitted $E_p$ (for D, E, and D+B+E in the left panel).}
    \label{fig:yonemati}
\end{figure*}

We tested the consistency of DBE with the known spectral-energy correlations of long GRBs: the $E_{p,z}$ - $E_{iso}$ Amati relation \citep{Amati2002}, which holds for time-integrated prompt spectra, and the $E_{p,z}$ - $L_{iso}$ Yonetoku relation \citep{Yonetoku2004}, which is valid for the peak prompt emission spectra. Starting from the results of our time-integrated analysis with the CPL model (Table~\ref{tab:spectral_fits}), we computed the isotropic equivalent energy, $E_{iso}$, and the rest-frame peak energy, $E_{p,z}$, for the three events, considering a range of redshift values $z \in [0.01,5]$. We also computed these quantities for the total duration of the event (denoted D+B+E), adopting as $E_p$ the most conservative fitted value, namely that of spectrum D. The results are shown in Fig.~\ref{fig:amati}, together with the $3\sigma$ scatter region around the Amati relation adopted from \citet{Nava2012}. To investigate the $E_{p,z}$ - $L_{iso}$ correlation, we extracted the 5-second peak spectrum obtained from the 5-second binning GBM light curve of each event, and fitted each of them with the CPL model to constrain the peak energy. We obtained $E_{\rm p} = 1.2_{-0.9}^{+2.5}~\rm MeV$ and
$F_{10~\rm keV-10~MeV} = (2.8_{-0.5}^{+0.3} \times 10^{-7})~\rm erg~cm^{-2}~s^{-1}$ for spectrum D, $E_{p} = 5.7_{-2.3}^{+4.2}~\rm MeV$ and $F_{10~\rm keV-10~MeV} = (5.6_{-1.5}^{+0.5} \times 10^{-7})~\rm erg~cm^{-2}~s^{-1}$ from spectrum B, and $E_{p} = 4.1_{-3.1}^{+15.3}~\rm MeV$ and
$F_{10~\rm keV-10~MeV} = (1.5_{-0.5}^{+0.2} \times 10^{-7})~\rm erg~cm^{-2}~s^{-1}$ for spectrum E, which corresponds to the brightest peak of the entire DBE event. Analogously, we computed the isotropic equivalent luminosity, $L_{iso}$, and $E_{p,z}$, for different redshifts. The results are shown in Fig. \ref{fig:yonetoku}, together with the $3\sigma$ scatter region around the Yonetoku relation adopted from \citet{Nava2012}. It is evident that the three events lie outside both the spectral-energy correlations that characterize long-duration GRBs.

\end{appendix}

\end{document}